# Exploring the Use of Enterprise 2.0 and Its Impact on Social Capital within a Large Organisation

Diana Wong, Rachelle Bosua, Sherah Kurnia, Shanton Chang
Department of Computing and Information Systems
The University of Melbourne
Victoria, Australia
Email: dianaw@student.unimelb.edu.au; rachelle.bosua@unimelb.edu.au

## Abstract

Despite the rampant adoption of Enterprise 2.0, there is lack of empirical evidence of how Enterprise 2.0 is aptly supporting the business objectives. Social capital theory will be used as a theoretical lens to understand the impact and implications of individual use of Enterprise 2.0. To ascertain the impact from the use of Enterprise 2.0 on the various dimensions of social capital, a single in-depth qualitative case study was conducted with a large professional services organisation. The findings unfold the different areas of impacts based on actual individual use and experience. The research concludes with a framework delineating the intertwined relationship between each social capital dimensions.

## 1 Introduction

Enterprise social networking site, or more commonly known as Enterprise 2.0, refers to a private, or internal social networking site for business use. Professor Andrew McAfee (2009) of Harvard Business School explained that Enterprise 2.0 is "the use of emergent social software platform that enables individuals to connect and collaborate through computer mediated communication and to form online communities." The central principles behind Enterprise 2.0, as emphasized by the author, are the simple free platforms for self-expression in publishing platforms, the emergent structure rather than imposed structure that enables new ideas to emerge organically, and the order from chaos that enable quick and easy filter of information. Enterprise 2.0 shares the convergence of Web 2.0 architecture (Levy 2009; McAfee 2006; Platt 2007) with the integration a wide range of social media tools which enables collective creation of web content and harnessing of users' collective intelligence (Soriano et al. 2008). A plethora of major social media tools, which are identified from the Web 2.0 platform, are such as blogs, microblogging, news feeds, wikis, social networks, tagging and social bookmarking. Enterprise 2.0 enables individuals to connect to each other and utilize the integrated social media tools in an activity stream that is relevant, secure, and collaborative (Sturdevant 2011).

Over the past years, the adoption of Enterprise 2.0 has gained significant momentum. Many of the leading professional services organisations, banking organisations, telecommunications organisation, government-linked and non-profit organisations are jumping on the bandwagon in adopting Enterprise 2.0 to promote communication and collaboration, with the aim of breaking down silos especially in large organisations. Nonetheless, according to (Chui et al. 2012), despite the numerous claimed benefits from the use of Enterprise 2.0, there are few organisations that are near to achieving the full potential benefit from the use of Enterprise 2.0. While there are wide spread interest and acceptance of Enterprise 2.0 within large organizations, there are still many wait-and-see attitude amongst business leaders to anticipate the purported benefits from the use of Enterprise 2.0. Hitherto, the potential impact of Enterprise 2.0 is still not well understood.

Meanwhile, there is also lack of empirical evidence of how Enterprise 2.0 is aptly supporting the business objectives. Up until now, although the use of Enterprise 2.0 has shown some business benefits, there are still many organisations that are sceptic towards the implementation and adoption of Enterprise 2.0. The negative implications of information security and privacy and loss of productivity seem to outweigh the potential benefits from the use of Enterprise 2.0. Organisations struggle to define the business case for implementing and adopting Enterprise 2.0 and are doubtful in the ability to capitalize on this social transformation. In addition, the direct return on investment from implementing Enterprise 2.0 remains unclear.

Through it all, what is really lacking amidst all the studies conducted is the underlying influence of knowledge sharing that enables the use of Enterprise 2.0 to truly impact business operations. Various authors have made claim of the increasing use of Enterprise 2.0 as an organisations' arsenal for knowledge sharing (Leidner et al., 2010; Levy, 2009; Li et al., 2012; Miles, 2009). The evolving landscape of knowledge sharing initiatives necessitate the leveraging of the emergent social networking technologies into a vehicle that is less top-down, corporate, monolithic, centric, database oriented, to an open, participative, people centric platform (McAfee 2006; Ribiere and Tuggle 2010).

The background and understanding of Enterprise 2.0 necessitates a theory-driven approach to understand the individual use and implications of Enterprise 2.0 in organisations. Based on the phenomenon of study, the research postulates the use of social capital theory which explains that the network connections between individuals within an organisational facilitate mutually benefit collective actions (Adler and Kwon 2002; Bourdieu 1986; Coleman 1988; Lin 1999; Nahapiet and Ghoshal 1998; Portes 1998; Putnam 1993; Uphoff 2000). Thus, to address the identified gaps, the central research question is as follow:

*How does the individual use of Enterprise 2.0 impact the social capital within a large organisation?*

To answer this question, the research studies the concept of social capital theory as the basis for describing and characterizing the organisation's set of relationships and Enterprise 2.0 use and intervention. An in-depth qualitative case study was conducted with a large professional services organisation, with existing implementation of Enterprise 2.0. Multiple data collection techniques were employed - in-depth interviews, documents analysis, and observation. The findings and discussions are presented in the last part of the research paper delineating the impact on various dimensions of the social capital.

## 2 Literature Review

Enterprise 2.0 is increasingly being adopted for knowledge sharing purpose (Bughin and Chui 2011; Levy 2009; Li et al. 2012; Miles 2009; Treem and Leonardi 2012). Riemer and Scifleet (2012) conducted a genre analysis on the use of Enterprise 2.0. The study explained that Enterprise 2.0 provides a common background through discussions and sharing updates between groups of individuals with similar interest. The emergence of shared background enables individuals to understand and interpret the problems, ideas, and knowledge shared. Other than that, the authors observed the common practice of posting contents, files, and links relates to sharing of knowledge. The action of sharing or posting contents, files or links lays the seed for other individuals to come across the information serendipitously and potentially get passed on to a wider social network.

Nonetheless, according to Majchrzak et al. (2013) the understanding of the individual use of Enterprise 2.0 should not be limited to the technology features, but should also focus on the action potential in the relationship between the individuals and the technology. The further explains the theory of affordance whereby the author outlines the relation between the environment and the object. In this context, it is crucial that we study the object which is the use of Enterprise 2.0 and how it affords social capital (environment). Past researches predominantly explored the relationship between personal social networking site – Facebook, and social capital among undergraduate students (Brandtzæg 2012; Ellison et al. 2007; Valenzuela et al. 2009). These studies have shown that social networking site is related to bonding ties such as reciprocity and trust and bridging ties such as spanning structural holes and network size. There are limited number of publications that exploring the use of social networking site and the implications on social capital. There are fewer publications geared towards Enterprise 2.0.

The study of the individual use of Enterprise 2.0 and its impact on social capital theory is important as the theory has been linked to the identification and positive implication to many organization benefits - breaking silos, improving collaboration, and improving knowledge sharing (Chiu et al. 2006; Huysman and Wulf 2005; Nahapiet and Ghoshal 1998; Wasko and Faraj 2005; Widén-Wulff and Ginman 2004). To further illustrate, social capital theory explains that social networks of different types and sources lead to organisation benefits (Adler and Kwon 2002; Bourdieu 1986; Coleman 1988; Lin 1999; Nahapiet and Ghoshal 1998; Portes 1998; Putnam 1993; Uphoff 2000). Bourdieu (1986), Coleman (1988) and Putnam (1993) define social capital as the aggregate of the actual or potential resources, which are linked to a network of relationships. Therefore, social capital constitutes a particular type of resource available to individual or a group in the social networks. The individual, for instance, owns some resources skills and tacit knowledge; other resources are not directly possessed and can be accessed within the social network connections. According to Yang et al. (2009) and Portes (1998) individual social capital refers to the resources generated by an individual's social network for his or her mutual benefit as a member of the network. On the collective or group level, social capital is often taken to be represented by norms, trust, and social cohesion (Putnam, 1993; Woolcock and Narayan, 2000). Nonetheless, the concept of social capital has expanded from individual to collective feature as Putnam (1993) argues that the social capital is about the connections among individuals where there arise the norms of reciprocity and trustworthiness. The result of subsequent studies (Adler and Kwon, 2002; Fukuyama, 2000; Nahapiet and Ghoshal, 1998; Woolcock and Narayan, 2000) support Putnam's assertion. According to a number of authors (Coleman, 1988; Markus and Robey, 1998; Nahapiet and Ghoshal, 1998; Yang et al., 2009;), social capital is still nonetheless concerned with both individuals and collectives. Hence, the mixing of the levels of analysis is useful in studying the impact of the individual use of Enterprise 2.0, since the impact is neither strictly collective nor individual in character.

Social capital theory is highly significant for the research on the use of Enterprise 2.0 as it addresses the mutually benefit collective social action which is facilitated by the various features of Enterprise 2.0. Over the past years, researchers have used the concept of social capital to explain that social networks of different types and sources lead to the sharing of knowledge (Nahapiet and Ghoshal, 1998). This perspective suggests that organisation enables social collectiveness (Kogut and Zander 1996) and knowledge sharing (Nonaka and Takeuchi 1995) whereby individuals mobilize available resources in order to contribute to collective goals (Spender 1996). As mentioned before, Enterprise 2.0 provides the mechanism to build social relationships, connect and collaborate, enables communication and ad-hoc sharing among individuals (McAfee 2009; Riemer and Scifleet 2012). While Enterprise 2.0 consists of a set of integrated social media tools, the use of Enterprise 2.0 can be classified into information sharing, collaboration and innovation, management activities and problem solving, communication, training and learning, and knowledge management (Turban et al. 2011).

The fundamental of Enterprise 2.0 is tightly coupled with social capital theory, which explains that social networks of different types and sources creates value for the organisation (Bourdieu 1986; Coleman 1988; Putnam 1993). Nonetheless, there is very limited number of publications exploring the use of Enterprise 2.0 and the implications on social capital. This poses the opportunity to explore the use of Enterprise 2.0 and its impact on the social capital, delineating how resources can be combined together through the use of Enterprise 2.0 to produce different outcomes and behaviours and to facilitate realisation of collective benefits. Therefore, this research adopts Nahapiet and Ghoshal's (1998) definition of social capital as the theoretical explanation on the interrelationship between the use of Enterprise 2.0 and social capital. The definition of social capital is similar to Bourdieu (1986), Coleman (1988), and Putnam (1993) whereby the authors view social capital as the sum of actual and potential resources embedded within, available through, and derived from the network of relationships possessed by an individual or social unit. Nahapiet and Ghoshal (1998) accounted for the influence of 'the structure and functioning of the social world' on the individual and provided the basic theoretical blocks to understand. Drawing on Nahapiet and Ghosal's (1998) research, the concept of social capital has great potential as an explanatory to the benefits of leveraging the use of Enterprise 2.0. The authors build on the existing theories and apply theories of individual motivations and social capital to develop a model for examining how individual motivations and social capital foster knowledge sharing. The authors have further categorized the understanding of social capital into the following three different dimensions:

- **Structural Dimension:** Structural dimension refers to the overall network configuration and network ties that contribute to cooperation, specifically mutually beneficial collective action, which is the stream of benefits that results from social capital. The network ties provide the channel for information transmission (Tsai and Ghoshal 1998) while the network configuration explains the underlying mechanism that leads to the emergence of social network ties. The varying strength of network ties creates opportunities for cooperation, trust, and empathy with others.

- **Cognitive Dimension**: Following that, the cognitive dimension refers to the ability to cognitively connect with each other in order to make sense of the knowledge context and communicate the meaning to others (Huysman and Wulf 2005). The higher the cognitive ability of individuals connected within a group, the better they communicate and share knowledge. The cognitive ability depends on many factors, which include the degree to which the individuals know each other, the topic of interaction, their familiarity with the topic of interaction or the extent to which they can anticipate the communication partner's current context of interaction.

- **Relational Dimension**: The structural and cognitive realms are essentially linked by the relational dimension. Trust, norms, obligations, reciprocity, and identifications are categorised as relational dimension which describes the process of linking and bonding individuals and groups (Chow and Chan 2008; Nahapiet and Ghoshal 1998). McFadyen and Cannella (2004) studied the strength of relational dimension that emphasizes the interpersonal exchange relationship. The authors emphasized that the relational dimension is developed through social interactions, which requires time, energy, and attention.

The dimensions of social capital have been found useful when exploring the use of Enterprise 2.0 and its influence on knowledge sharing. It is mentioned that social and information phenomena are deeply anchored in each, therefore social capital theory is cited often as a suitable theoretical framework to explain knowledge sharing mechanism in organisations (Hansen et al. 1999; Nahapiet and Ghoshal 1998). Therefore, to ascertain the impact from the use of Enterprise 2.0 on the various dimensions of social capital, an in-depth case study was conducted and is explained in the following section.

# 3   Research Method

A qualitative case study research method was selected as the approach to acquire a deeper understanding of the individual use of Enterprise 2.0 and how it impacts the social capital whilst unfolding the behaviours, perspectives, feelings, and experiences of individuals using Enterprise 2.0 (Benbasat et al. 1987; Yin 2013). This entails employing multiple methods of data collection which includes observations, semi-structured interviews, and documentation analysis. The in-depth case study was conducted with a single large professional organisation that has multiple offices worldwide and within Australia. Twelve participants across the organisation were recruited for the case study – 2 Partners, 4 Directors, 1 Manager, 1 Yammer Community Manager, 2 Senior Consultants, and 2 Analysts. Purposeful sampling was used for the recruitment of interview participants from varying roles and responsibilities to yield richer and all-inclusive view of the case study. The concept of saturation applied, as the researchers concluded data collection following the 12th participant as there is no more new information or themes emerged after interviewing the last participant. As mentioned before, the unit of analysis is on the individual use of Enterprise 2.0 whilst examining the impact on both individual and collective social capital. The understanding on the use of social resources requires a more complex context to understand the access to social resource and the other collective aspects of social capital such as norms, trust, obligations, and identification. The composition of the interview participants consisted of individuals with experience in using Enterprise 2.0, and who used it on a regular basis (on average 3 times a week). Their experience of using Enterprise 2.0 was based on (Steinfield et al. 2009) research, using a six-week benchmark recruiting individuals who had joined the organisation at least 6 weeks prior, would have had an opportunity to use the Enterprise 2.0 and received potential benefits from using Enterprise 2.0. This was to ensure adequacy and relevancy of information gathered to examine the impact from the use of Enterprise 2.0 on social capital. The individual interviews lasted an average of 60 minutes each. The interview protocol consisted of approximately 10 open-ended questions inquiring participants of their use of Enterprise 2.0. Data gathered from the case organisation is analysed using thematic analysis as suggested by (Miles and Huberman 1994). The transcription of the interviews, documents, and notes from the observation was analysed and involved identifying initial codes, co-occurrence codes, and graphically displaying the relationships between codes within the data set. The process leads to identification of themes classes and subclasses through careful reading and re-reading of the data as recommended by (Klein and Myers 2001).

## 3.1   Background of Case Organisation

The case organisation is a large professional services organisation that has multiple offices worldwide and within Australia. The organisation adopted Yammer – major vendor of Enterprise 2.0, as its enterprise social networking platform back in September, 2008. The Yammer adoption begins from the grass roots, innovators and early adopters – the perceived 'geeky types', with their strong technology orientation, first use Yammer based on their perceived needs. Nonetheless, the Chief Executive Officer of the organisation was very supportive of innovative ways of improving business and productivity. By April 2009, there were already a couple of hundreds of employees who have joined Yammer within the organisation. The participation increased exponentially during the logo campaign. The management wanted a way to engage employees to submit key words which signified the organisation's strategy. Yammer became the solution to reach out to the broader audience. The logo campaign group was set up in Yammer, and an email was circulated inviting employees to join in Yammer and participate in the logo campaign by submitting a tag line suitable for the organisation's logo and brand. Within minutes, the management witnessed a drastic increase in employees joining Yammer. This was when Yammer gained real momentum within the organisation. Data captured in April, 2011 shows that 5124 users joined the network with well over 4000 active members who also created 694 groups.

# 4   Findings and Discussion

## 4.1   Broadens Social Network

Based on the findings, all the interview participants responded that the use of Yammer broadens their existing social network. The interview participants responded that they follow and interact with other employees for various reason - interesting posts, and keeping in contact, regardless of whether they know or did not know the person, or if they were from different departments or hierarchy in the organisation. As some interview participants described:

*"The use of Yammer flattens the organisation, so anyone can associate and socially respond equally well, if a partner posts a question, or a graduate post a question." – Partner*

*"With Yammer, you're expanding your network exponentially because there are people that know people that you have no idea who they are, who respond to you. So that's how you're actually broadening your network." – Partner*

From social capital's perspective, the broadening of social network involves the configuration of social network and can be further examined by the strength of the social network ties. In Putnam's view, bridging social capital occurs when there is formation of weak ties. This is apparent in the findings whereby individuals are connected with other individuals outside of their existing social network.

*"It deepens your existing relationship, it widens your network, and I constantly add people to my network. A lot of those people are people I know outside work or at work, but there are people I made literally on Yammer, I never met them face-to-face. Because we are collaborating so much, I feel I know them really well." – Yammer Community Manager*

## 4.2   Establishes Serendipitous Connection

The social capital is augmented whereby the use of Yammer supports loose social ties, allowing individuals to create and maintain larger diffuse networks of relationships from which they could potentially draw resources. (Granovetter 1973) classic work on the strength of weak ties explains that these loose social ties present opportunities for bridging new social networks and increase the efficiency of information diffusion and are more likely to be the source of innovation. The findings explain these loose ties as 'Super Connectors' or 'Navigators'.

*"You're expanding your network exponentially because there are people that know people that don't know people that you have no idea who they are, that respond to you." – Director*

*"Because I actually meet people like, basically there are people who have wide networks. So if you want to know some detail about something, it's best to reach out to them because they then get probably in touch with people in different places who could help you. It is just good way to know who is the best person to get in touch with." – Analyst*

The use of Yammer also helps reduces pain points by connecting networks of individuals who can solve problems or help navigate towards solutions in real time. On this notion, most interview participants would use Yammer for crowd sourcing where one of them quoted a post:

*"I have a client that has a particular problem, does anyone know something about whether this industry or something". – Senior Consultant*

Nonetheless, individuals can also be seen using Yammer to connect with other individuals of the same workgroup or common interest. (Granovetter 1973) described strong network ties as individuals who share the same associations or groups. They tend to form dense clusters of strong ties, and they are more likely to maintain a reciprocal communication with each other through often responding to comments and postings. Individuals are dedicated to keep the discussion going and garner participation from one another. From one of the excerpts of the interviews:

*"People that work across the industry like financial services that might work for different client but they collaborate across the industry within Yammer group." – Partner*

The stronger the network ties, the more similar they are (Granovetter 1973; Putnam 1993). Hence, it is often that the knowledge shared within the groups is homogenous across the topics of discussion. From the findings, groups founded on key organisation initiatives have a dense network based on the active postings and comments from group members. Campaigns and project groups too result in creation of a dense network either by obligated participation from project members or intrinsic motivation for campaigns.

### 4.3 Breaking Down the Wall of Communication: Functional and Hierarchical Barriers

Subsequently, the use of Yammer enables employees to reach out to other employees across functional groups. The use of Yammer reaches out to all levels of employees, allowing employees to follow and connect with, follow and be followed by any employees within the organisation. Therefore, the use of Yammer impacts the social capital such that it breaks down functional silos and hierarchical barriers. With Yammer in place, individuals are able to connect with the whole organisation having access to faster and wider information and resources. Social network ties are formed far more easily than in the past. As mentioned by one interviewee:

*"I think generally people are more connected through that, because you tend to see this little face next to everyone's face. You actually meet people across the business. You often communicate with someone via Yammer and you never actually see them face-to-face and when you actually do meet them you know more about them because you have been sharing." – Senior Consultant*

The use of Yammer drives the notion of meritocracy flat. Many of the interview participants responded that the comments on the Yammer postings would come from anyone including the Chief Executive Officer or senior leadership team.

*"What Yammer does is, it smashes hierarchies because the culture here is, I know a lot of people here, but when you put 6000 people here in the organization, everyone knows who the top guys are, you don't say before I answer this post, is this person the same level with me. It doesn't matter if they are senior or junior." – Partner*

*"Because it is a fairly flat structure by Yammer, you communicate with people across different levels and you communicate with people outside the core competency. We have seen the degree of connections in our organization." – Senior Consultant*

### 4.4 Connects Individuals of Similar Experience and Interest

The findings demonstrate that the use of Yammer enables employees to connect with networks of other employees who share similar experience. Interactions between employees are framed by a unique common context. Employees who encounter the similar problem in the past are able to interpret the situation and may participate in the thread of discussion and help to navigate to a solution.

*"Sarah talks a lot about this particular tax issue, next time I need something, if I don't know who and I post to your company feed. And If I know something then I might go, add Sarah to see if Sarah can help." - Director*

One of the interview participants use Yammer for crowd sourcing as ways to de-risk the project, asking questions or information at the start of the project from the network if anyone has done something similar for the project that can be reused or referenced. She highlighted an example whereby she needed to put together a report for a banking client to envisage what banking of the future would look like if the centre of banking shifted from London and New York to Shanghai or Beijing. She then posted on the all company news feed:

*"What would happen if the whole centre of gravity shifts from the West to the East? What would change?" – Director*

As a result, the wisdom of crowd generated stream of ideas from employees who shares similar experience and interest in the topic, which she then gathered together and presented to the client.

### 4.5 Creates Awareness

The use of Yammer allows categorization and identification of common interest groups, which creates the awareness of where knowledge resides. One of the interview participants explained how the project team members uses Yammer and to provide update and information on the project to keep everyone across on the progress of the project. In his response, he mentioned:

*"There is other times we might be doing something like a Digital strategy, we might use Yammer as a way to share everything from the minutes, the agenda of the meeting, the minutes of what happened, the activities, the photos, the whiteboard, the work that clients they go "Just heard about something, here's the PDF what do you think it is", or just read this Gardner report or something. So, it can be really quite open." - Partner*

Yammer is also commonly used as a platform to gather the thought of employees whilst driving and promoting a campaign to create awareness. Interview participants mentioned that the use of Yammer enables employees to keep abreast of the current news and happenings within the organization.

*"We have got a team in Adelaide, Melbourne, and Sydney. They can all see what is going on." – Senior Consultant*

Within the same context, the use of Yammer increases individuals' awareness and understanding of the work group, organisation, and business, leading to serendipitous opportunity.

*"So, I have got on some projects because I have been active on Yammer. I worked on some project with them, word out that they need people with this background, so, I put on my hand up." – Analyst*

Other than that, employees are able to identify and locate where knowledge resides through common interest group. The level of understanding and the ability to cognitively connect with each other is dependent on the level of association with the individuals or groups.

*"I belong to quite a significant number of groups. I wouldn't say I read everything on the group. I kind of monitor, on the left side you can see all the new postings in each group. But I won't necessary click on the new postings in each of those groups. It will only be the ones I am really connected with and some of them, for example, if someone posts into Yammer Crowd Support, I actually get an email notification." – Yammer Community Manager*

### 4.6 Establishes Shared Understanding

As one of the partners responded, when projects are managed via Yammer, the whole project team can see everything that is going on, not just the deliverables but also other project artefacts, the conversations of what happened during the meeting with client or if they have any concerns or issues. As individual user, if he or she reads through the history, they can get a much better perspective about what was going in the project. On a different note, when someone leaves the project team and a new person joins the project team, 3 months into the 6 months project. They can very quickly look into the history because all the communications for the project is within Yammer and not scattered across emails. As one of the interview participants responded:

*"Yammer is actually acting like what they used to call the water cooler or the photocopier, where you actually gather and have chat and share information that makes you more familiar with the company and the rest of the company more familiar with you." – Director*

### 4.7 Fosters Norms of Behaviour

As gathered from the case study, culture, is the key to achieving the critical mass from the use of Enterprise 2.0. The organisation has strong culture embedded among the employees. The CEO described that: *"I believe culture eats strategy for breakfast. We've seen and heard from many leaders that even with a good strategy, we may not perform well if we don't have a culture of executing the strategy."*

This is as reflected in the interview participants' use of Yammer, with the tone of voice, the conversational postings, with occasional humour injected in the threads of discussion. The senior management involvement is pivotal in driving this norm. As highlighted in one of the interviews, there was an event whereby some user mistakenly entered the letter 'A' to the all company news feed. Consequently, the post got published in the all company news feed. Some other user thought it was funny and started the letter 'B', and it may not have been a letter 'B', but it might have been a picture of 'B'. Since then, it continued and someone else sent it a 'C' and it went on and on. Recently, as one of the interviewee mentioned, some user started the thread again.

*"So we have gone through the alphabets, it kind of is a way of creativity. It is the community having fun. Imagine our company, we have a mixture of a whole lot of different personalities; it is quite a different culture there. It is really nice that people are having a little bit of fun in there." - Yammer Administrator*

The post actually sets a real tone - a conversational cross hierarchy tone. It is by means of injecting fun through informal communication about clearly non-work related matters. More importantly, the Chief Executive Officer and the senior management encourage this notion as this drives a very open environment within Yammer. The use of Yammer promotes norms of social behaviour, which is essentially a reflection of the organization culture. Employees are welcomed to express their concerns and opinions as long as it is appropriate within the context of the organisation, share some humour and socialize with other employees. It is reported in the findings that although there isn't formal training provided on the use of Yammer, new employees on the network are able to pick up the norms based on observing how other individuals in the social network use Yammer.

### 4.8 Builds Trust and Empowerment

Besides, the use of Yammer is also about building trust and empowering employees. Most interview participants expressed that they generally feel safe using Yammer as the organisation provides an open and transparent environment. As described by some interview participants, it is safe to assume that most employees would know the appropriate content to discuss openly in Yammer.

*"Similar to email or offline conversations, you would not send out or comment on anything that is generally not acceptable within an organisation." - Analyst*

*"Because it is empowering trust, other people just say 'hey, that is ridiculous thing to say in Yammer.' They are pretty open, pretty open culture." - Yammer Community Manager*

There were few occurrences of controversial topics being discussed in Yammer. One example provided was the thread of discussion on the change in the expense claim policy. The change in the policy has caused an unhappy employee to express the dismay of the new claiming process. There were series of comments from other users, and this caught on the attention from the Financial Controller who implemented the change. This situation was not seen as a taboo in the use of Yammer within the organisation. On the contrary, the CEO commented that this was a good and healthy discussion. It provides an opportunity to resolve an issue. The senior management looked through the discussion and worked with the employee who initiated the post. The issue got resolved, and the expense claim system was revised to the original method. The example provided in the threads of discussion drives the notion of meritocracy flat. Employees can trust the social network whereby their opinions are valued and seen as an opportunity to learn and improve.

*"Within yammer, people can draw upon the collective knowledge of the firm in a very informal and non-threatening way." – Senior Consultant*

It is also observed that individuals do not believe in hoarding knowledge for power and draw on the collective knowledge of the organisation. This is apparent in many of the examples where individuals crowd source for reference of previous project work done or any other information. Individuals were subsequently asked if they trust the information provided whether it is reliable and accurate. The responses gathered were in unison that they would trust the information provided. Individuals generally would not post information that is not reliable as it is opened to the network. The information provided will be crowd checked and self-correct if it is not accurate. In circumstances that the individual is uncertain of the information, there are other method to assess the validity or accuracy of information such as looking up the individual's past postings and comments, the service line they are involved in and their areas of interest. In anyways, as mentioned by one of the interviewees, trust is builds over time through engagement, and that is when trust will also increase in time.

*"All of us are smarter than one of us. If you are always looking to tap into <the crowd>, you are able to bring that knowledge to sow a solution to your customer at the faster pace of time, which is more successful." - Partner*

### 4.9 Develops Understanding of Interpersonal Obligations

Other than that, the interview findings present mixed of responses in terms of obligation to respond to the network or groups. Employees generally feel obligated to respond to the groups which they feel a great sense of belonging or when they are tagged in a post. There are also employees who do not feel obligated to respond to the discussions or postings.

*"I try to help people when they try to find something, answer questions about Yammer functionality, or post to the Yammer customer network to try and get help there." – Yammer Community Manager*

One of the very interesting finding was the feeling of obligation also comes in the form of voluntary social engagement. Through the use of Yammer, employees are intrinsically driven and excited to participate voluntarily, taking time outside of work responsibilities to contribute to the organisation. The organisation set up a reverse mentoring program for older employees who weren't used to using social media to learn and adopt Yammer. For example, if the employee is a 'Digital Dinosaur', s/he can be adopted by a 'Digital Native' and they will turn them into a 'Digital Immigrant'. Some of the interview participants expressed that they feel passionate and volunteer to take ownership of the initiative. One of the interview participants described a posting on Yammer:

*"All you Digital Natives in Melbourne, we need your help out there. There is a movement in Jurassic Park. There are 50 Digital Dinosaurs has signed up to receive your expert tuition to become Digital Immigrants. Let me know if you want to participate."- Partner*

This initiative can be viewed both ways, as the Digital Natives wanting to innovate how things work in the organisation, or the older generation employees who are driven and want to encourage younger employees to get involved. Both sets of user volunteer to keep the momentum and critical mass going in Yammer. As a final point, reciprocity plays a part in ensuring continuous social interaction between individuals. The findings explained that often when an individual user stops giving back or contribute to the discussion, other individuals in the network will feel less obligated to reply. One of the interview participants responded that reciprocity is like social currency.

*"If you share interesting knowledge or good knowledge, people are more inclined to reciprocate." - Analyst*

The findings explained that often when an individual user stops giving back or contribute to the discussion, other individuals in the network will feel less obligated to reply.

## 5   CONCLUSION

Finally, the understanding of the use of Enterprise 2.0 and its impact on social capital can be summarized in Figure 1.

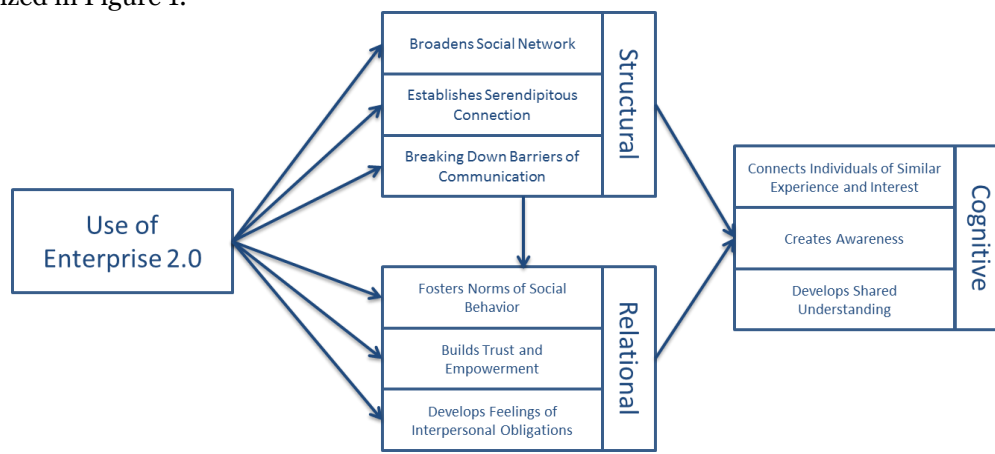

*Figure 1: Framework Depicting the Use of Enterprise 2.0 and its Impact on Social Capital*

To sum up the finding and discussion, the network of an organisation is formed through vertical and lateral association of individuals and groups. The use of Enterprise 2.0 facilitates the formation and varying strength of the social network ties creates opportunities, provide accessibility, and novelty of knowledge across the network of individuals, positively impacting the social capital within the organisation. This is evident from the findings that the use of Enterprise 2.0 impacts the social capital such that it broadens the social network, establishes serendipitous connection between individuals - bridging and linking, and breaks down the barriers of communication. The use of Enterprise 2.0 also impacts the relational dimension whereby it fosters and deepens relationships through social interactions within Enterprise 2.0. As evident, the use of Enterprise 2.0 shapes the norms of behavior through observing other individual's and the continuous social interaction that occurred over the network. The findings also confirm the impact on trust whereby individuals share their opinions and thoughts, ideas, problems with the social network without fear of being reprimanded or used against them. The interview findings present mixed of responses in terms of the impact to obligation to respond to the network or groups. Individuals generally feel obligated to respond to the groups which they feel a great sense of belonging or when they are tagged in a post. Similarly, the strength of the network ties and reciprocity plays a part in influencing obligations. This leads to the relationship between structural and relational dimension.

The model illustrates that there is an impact of structural dimension to relational dimension whereby the strength of the formal and informal relationships between individuals within the network impacts the social inclusion, which translates to relational dimension. Meanwhile, the impact on cognitive dimension depends on factors, which include the degree to which the individuals know each other, the topic of interaction, their familiarity with the topic of interaction or the extent to which they can anticipate the communication partner's current context of interaction. This is reflected in the findings based on the shared understanding, awareness and sharing of similar experience and topics of interest. The impact on cognitive dimension also stems from activities such as crowd sourcing for managing exceptions (problem solving), seeking opportunities or brainstorming of ideas. Finally, coherent with the Vandaie (2007), the cognitive dimension is shaped by the joint influence of structural and relational dimension. The use of Enterprise 2.0 and its impacts on the structural and relational dimension explains the formation of attitudes, perceptions, and beliefs about the organizational environment.

In conclusion, the use of Enterprise 2.0 provides the avenue for social networking and interaction, which fosters strengthening of relationships between individuals, leading to positive impact on the social capital dimensions. The characteristics and use of Enterprise 2.0 poses great ability to influencing collective benefits based on the impact to the dimensions of social capital. However, further investigation is required to obtain an understanding the interdependencies between each dimension in Figure 1 and if there will be any variation from the individual use of Enterprise 2.0 across different industry sectors, and if it impacts the social capital in the same way. Through this research, it is hoped that the empirical evidences discussed will provide a stronger business case for organizations to implement and encourage and the use of Enterprise 2.0 towards achieving competitiveness.

## Acknowledgements

I am thankful to my supervisors for their guidance and motivation. I would also like to acknowledge with deepest gratitude, the support and love of my family and close friends.

## Copyright